# The Four-Step Action Loop with Feedback: A teachable perception–action framework for everyday self-regulation

Sundeep Kapila[a,b]

[a] The Swasth Institute, a program of Swasth Foundation, Mumbai, India

[b] Centre for Traditional Indian Knowledge and Skills, Indian Institute of Technology Bombay, Mumbai, India

Correspondence: sundeep.kapila@swasth.org · ORCID: https://orcid.org/0009-0004-0670-142X

 

## Abstract

Everyday stress is largely a learned pattern: a familiar cue, a fast reading of threat, an automatic reaction and a body that stays activated. Body-based practices such as mindfulness, interoceptive training, Somatic Experiencing and yoga target these patterns, but they are usually taught as techniques without a map of the mechanism they act on. Pain neuroscience education suggests that understanding such a mechanism can itself change experience. No comparable model exists for everyday self-regulation. This paper proposes the Four-Step Action Loop with Feedback, developed over six cohorts of a workshop for non-clinical adults. The loop describes any action as four steps: Sense (exteroception, proprioception and interoception), Process (perception as inference, weighted by context and safety), Choose (habitual reaction or goal-directed response after a pause) and Execute (autonomic and somatic output). Feedback from each outcome acts beneath all four steps, through synaptic plasticity, habit formation and allostatic load. The paper integrates the perception–action cycle, information-processing stages of choice, interoceptive inference, the autonomic space model and the defence cascade. It places a set of practices on the loop: full sensory awareness, grounding, orientation, resourcing, the pause, reduced stimulus and choice, slow movement and relaxation, each with a proposed mechanism and an assessment of the evidence. It also describes the model's translation into four practitioner skills. Seven testable propositions are derived for future evaluation. The model is offered as a teachable framework for wellbeing, not a treatment.

## 1. Introduction

Most stress is not a response to present danger. It is a learned pattern: a familiar cue, a fast reading of threat, an automatic reaction, and a body that stays activated long after the moment has passed. People often understand this in principle and still find the pattern repeating. Insight alone rarely changes a reaction that runs below deliberate choice.

A growing range of practices targets these patterns from the body upward. Mindfulness trains non-judgmental attention and decentering (Bishop et al., 2004). Interoceptive training builds awareness of internal bodily signals (Price & Hooven, 2018). Somatic Experiencing works with sensation and the completion of defensive responses (Payne et al., 2015). Yoga combines posture, breath and attention (Gard et al., 2014). Each has its own vocabulary, and each is usually taught as a set of techniques. What is rarely given to the learner is a map of the mechanism: an account of what happens between a cue and a reaction, and of where in that sequence a given practice acts.

There is evidence that such a map can matter. Pain neuroscience education teaches people with chronic pain how the nervous system produces pain. Learning the mechanism has been found to reduce catastrophising and fear of movement (Moseley & Butler, 2015). Understanding the mechanism can itself change experience. No comparable teaching model exists for everyday stress and self-regulation in non-clinical adults.

This paper proposes one. The Four-Step Action Loop with Feedback describes any action as four steps (Sense, Process, Choose and Execute), with feedback from each outcome acting beneath all four. The model has been developed and taught over six cohorts of a five-session workshop and has been extended into a certificate course for practitioners.

The paper makes three contributions. First, it brings together established accounts (the perception–action cycle, the information-processing stages of choice, interoceptive inference, autonomic regulation and experience-dependent plasticity) into one loop that a non-specialist can use. Second, it places a set of practices on the loop, so that each has a stated target and a proposed mechanism. Third, it derives seven testable propositions. Section 2 describes the approach. Section 3 sets out the loop, Section 4 the practices, and Section 5 the propositions. Section 6 relates the model to existing approaches, and Section 7 discusses safety and limits.

## 2. Approach

This is a conceptual paper. It follows the theory-synthesis approach described by Jaakkola (2020), which builds a framework by integrating existing theories around a phenomenon, here the sequence from cue to action and its repetition. Sources were selected for their relevance to a step of the loop and their standing in the literature. Where the model relies on a clinical hypothesis with limited evidence, such as the completion of defensive responses, this is stated.

The model was developed through practice before it was written down. It began as a workshop on the biology of the cell, the neuron and the autonomic nervous system, with no organising model. Over the following cohorts, a reaction-versus-response distinction was added, then a stimulus–action box that was gradually opened into separate steps. By the fifth cohort, in 2025, the full sequence of steps with feedback was taught from the first session, with one session devoted to each step. Participant requests for a single framework that runs through the whole course contributed to this change. These observations shaped the model. They are not offered as evidence for it, and an evaluation of the workshop is planned separately.

The author's background also shaped the model. He trained in computer science and has spent about 25 years working on how information is gathered, processed and acted on: in software development, which he has continued throughout, in management consulting and in public health. The loop's input–process–output form reflects that information-processing view. He is a certified coach and a trained Somatic Experiencing practitioner and yoga therapist, and his doctoral work modelled transport processes including neurotransmitter release (Kapila & Nair, 2025). The model therefore draws on engineering, contemplative and somatic traditions as well as neuroscience. The paper aims to keep its claims within what the neuroscience literature supports.

## 3. The Four-Step Action Loop with Feedback

The model describes any action, from a reflex to a considered decision, as four steps in sequence: Sense, Process, Choose and Execute (Figure 1). A fifth element, Feedback, is not a further step. It is a layer beneath all four. The outcome of each action, and the bodily state it leaves behind, changes how the next round of sensing, processing, choosing and executing will run. This is how a single reaction becomes a pattern.

The loop takes its outline from three established accounts. The first is the perception–action cycle, in which sensory and motor hierarchies are linked by continuous loops (Fuster, 2004). The second is the information-processing account of choice, which divides the interval between stimulus and response into identification, selection and programming stages (Schmidt et al., 2019). The third is predictive accounts of the body, in which perception and action are both shaped by the brain's expectations about bodily states (Seth, 2013; Barrett & Simmons, 2015). The model does not add new mechanisms to these accounts. Its contribution is to arrange them so that a non-specialist can locate their own experience on the loop, and so that each practice can be tied to the step it is meant to change.

Each subsection below defines one step, summarises the evidence on which it rests, and names the practices that act there. Section 4 then sets those practices side by side.

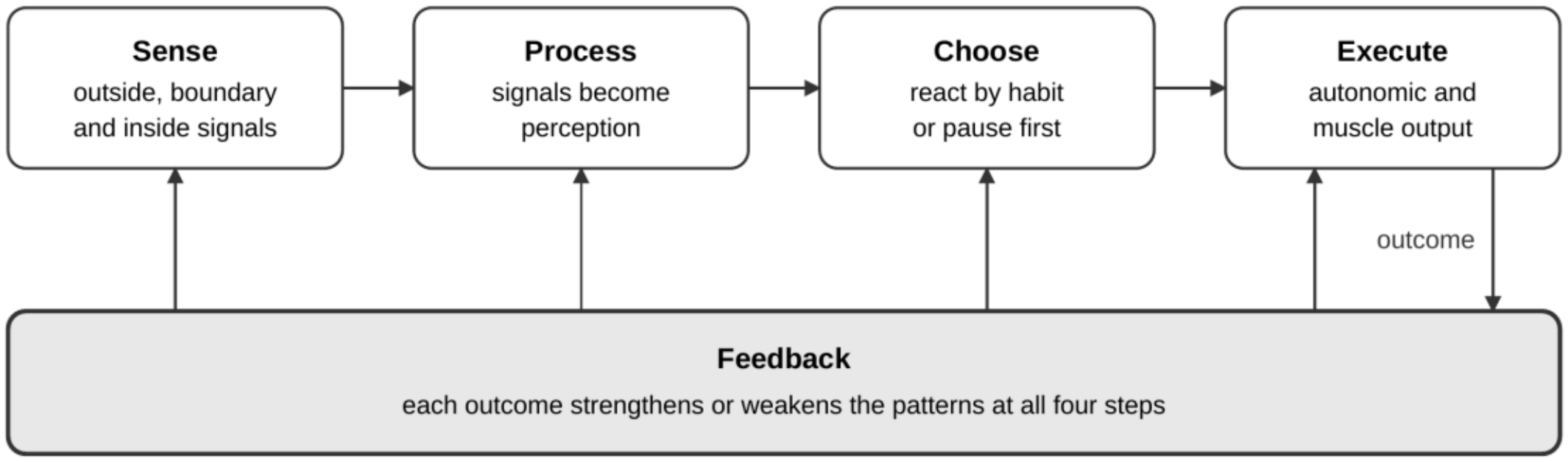

**Figure 1.** The Four-Step Action Loop with Feedback. Sense, Process, Choose and Execute run in sequence within one action. The outcome of each action feeds back to all four steps, changing how the next action will run.

### *3.1 Sense*

Every action begins with signals from receptors. The model groups them into three channels. Exteroception carries information about the world: sight, sound, smell, taste, touch and temperature. Proprioception, with the vestibular sense, carries the position, movement and load of the body. Interoception carries the state of the body's interior, including heart, lungs, gut, temperature, pain and chemical balance (Craig, 2002; Khalsa et al., 2018). For teaching, the three are named as outside, boundary and inside.

Interoception is the least familiar of the three to most adults, and the most relevant to stress. Its signals are not limited to the gut. Baroreceptors in the major arteries report each heartbeat, and the timing of the cardiac cycle changes how threat-related stimuli are processed (Garfinkel & Critchley, 2016). Interoceptive ability is also not one thing. People differ in how accurately they detect signals such as their heartbeat, in how confident they are in those judgements, and in how far the two agree (Garfinkel et al., 2015). A mismatch between predicted and actual body states has been proposed as a basis of anxiety (Paulus & Stein, 2010).

This matters for the loop because the Sense step sets what is available to everything after it. A signal that is missed cannot be weighed. A noisy one can be read as more alarming than it is. Practices aimed at Sense therefore try to widen attention across all three channels and to make interoceptive signals more reliable. Somatic Experiencing (SE) makes this its starting point, directing attention to visceral and musculoskeletal sensation before thought or emotion (Payne et al., 2015). The workshop practice of full sensory awareness, a slow scan from the feet to the room, follows the same bottom-up order.

### *3.2 Process*

Signals do not arrive as experience. They are turned into a perception or feeling by comparison with what the brain expects. In predictive accounts, the brain continuously predicts its sensory input and updates on the difference, weighting each error by how reliable it seems (Friston, 2010). Applied to the body, this means that a feeling such as anxiety or calm is an inference about bodily state, not a direct readout of it (Seth, 2013; Barrett & Simmons, 2015).

Two consequences follow. First, the same signal can carry different meanings in different contexts. A racing heart and sweating palms are read as threat before a difficult conversation and as effort during a run. Second, body signals and outside cues are weighed against each other. Balar and Kapila (2025) proposed that this balance can be described as a weight shifting between the interoceptive and exteroceptive streams, and that anxiety involves too much weight on bodily threat signals. Whichever formal account is preferred, the practical point is the same: changing the context, or the cues the brain attends to, can change what a body state means.

This gives safety a central place at the Process step. The generalized unsafety theory of stress argues that the default state of the stress system is threat, and that the stress response persists until safety is perceived (Brosschot et al., 2016, 2018). On this view, much chronic stress reflects a failure to register safety rather than the presence of danger. SE practices of resourcing and orienting act here. Resourcing brings attention to a remembered or present source of safety. Orienting turns the head and eyes to scan the room, giving the brain direct outside evidence that it is safe (Payne et al., 2015).

The Process step also has limited capacity. Neurons code stimulus intensity largely by firing rate and by how many neurons are recruited (Kandel et al., 2021). Sustained high input has a metabolic cost, and repeated activation of the same pathways lowers their threshold. The model teaches four ways to reduce the load: less stimulus, lower intensity, higher threshold through regulation, and fewer choices. The last of these connects to the next step.

### *3.3 Choose*

Once a situation has been perceived, an action must be selected. The model distinguishes two routes. A reaction is fast and automatic: a learned pattern runs without deliberate choice. A response follows a pause, in which the person notices the state they are in and selects an action that fits the present situation. Reactions are not treated as faults. Patterns form because they save time and effort: a well-learned action runs faster and with less deliberation than one chosen afresh each time (Wood & Rünger, 2016). Most daily behaviour runs on them, and they free capacity for what matters. The aim is to make a pause available when a pattern no longer serves, either because the context has changed or because its outcome has become costly.

Two examples show how a useful pattern can become costly. A child who learned to stay quiet around a volatile adult was protected by that pattern. As an adult, the same silence may run in meetings where speaking up is safe and needed, and colleagues may read it as a lack of ideas or competence. Checking the phone at every alert once brought timely news. Repeated a hundred times a day, it fragments attention and keeps the body mildly activated. In both cases the pattern was a sound solution to an earlier situation, and the cost appears only when the situation changes.

The distinction maps onto the well-established contrast between habitual and goal-directed control. Habits are triggered by context and run with little reference to their current outcome. Goal-directed actions are chosen with the outcome in view, and draw on prefrontal control (Dolan & Dayan, 2013; Wood & Rünger, 2016). Stress shifts the balance towards habit (Schwabe & Wolf, 2011), which is one reason people under pressure repeat responses they know are unhelpful. The neurovisceral integration model links this to the body: prefrontal regulation of behaviour is coupled to cardiac vagal control, indexed by heart rate variability (Thayer & Lane, 2000).

The information-processing account explains why the pause takes time and effort. Between stimulus and response lie stimulus identification, response selection and response programming (Schmidt et al., 2019). Selection time rises with the number of options. Merkel (1885) first showed this, and Hick (1952) and Hyman (1953) formalised it: choice reaction time grows with the logarithm of the number of equally likely alternatives. Each doubling of options adds a roughly constant delay. Fewer choices therefore leave more capacity for a considered response, which is why the model treats simplifying routines as a regulation practice and not only a convenience.

Pausing also changes how a thought or feeling is held. Mindfulness research describes this as decentering: experiencing a thought or sensation as a passing event rather than as a fact about oneself (Bishop et al., 2004; Bernstein et al., 2015). Decentering is a plausible mechanism by which the pause in the loop weakens automatic reactions.

### *3.4 Execute*

All action is carried out by muscle, and the model divides execution accordingly. Involuntary execution runs through smooth and cardiac muscle and glands under the autonomic nervous system: heart rate, breathing, digestion and immune activity. Voluntary execution runs through skeletal muscle under the somatic nervous system: posture, movement and speech. A stress reaction is therefore not only a thought. It is an autonomic and muscular event, and it can be worked with at that level.

**Autonomic execution.** The workshop image is of an accelerator and a brake. The sympathetic branch mobilises energy for action; the parasympathetic branch supports rest, digestion and repair. A common simplification places the two at opposite ends of one scale. The autonomic space model shows instead that the branches can vary independently: both can be low, one can dominate, or both can be high together (Berntson et al., 1991). This captures the state often described as feeling "stuck": a state of high sympathetic drive held in check by strong parasympathetic restraint, felt as frozen or shut down. It also explains why the goal of regulation is flexibility, a smooth return to baseline after activation, rather than permanent calm. The paper uses this model rather than the polyvagal hierarchy, whose anatomical and evolutionary claims are disputed.

**Defensive responses.** Under threat, execution follows a graded sequence that has been described across species: arousal and orienting, then fight or flight, then freezing, and finally collapse when escape seems impossible (Kozlowska et al., 2015). Each is an adaptive programme, not a failure. SE adds a clinical hypothesis: when a defensive response is mobilised but cannot be completed, the arousal that drove it may persist, and bodily symptoms and stuck patterns follow (Payne et al., 2015). Gently completing the interrupted movement, slowly and with attention, is proposed to let that arousal settle. SE as a whole has preliminary support: a scoping review found 16 studies, only two of them randomised trials (Kuhfuß et al., 2021; Brom et al., 2017; Andersen et al., 2017). The completion mechanism itself has not been tested directly.

**Somatic execution.** Skeletal muscles can only pull. They work in opposing pairs, so a movement is efficient only when the opposing muscle releases. The smallest unit of control is the motor unit, one motor neuron and the fibres it drives (Kandel et al., 2021). Two practical points follow. Chronic tension in antagonists makes movement effortful and reduces usable strength. Low-force, slow contractions recruit the smallest motor units first, so force can be graded in finer steps (Henneman, 1957; Henneman et al., 1965). Slow, attentive movement methods such as Feldenkrais improved balance in older adults in pooled trials, although risk of bias was high and evidence for wider benefits rests on single studies (Hillier & Worley, 2015). The workshop summarises this as "slow is fast", treating slow, attentive practice as the route to fine control. The same principle appears in an older contemplative source. The Yoga Sūtras define posture as steady and comfortable (*sthira-sukham āsanam*, YS 2.46). They describe it as perfected through relaxation of effort (*prayatna-śaithilya*) together with absorption of attention (*ananta-samāpatti*,

YS 2.47) (Woods, 1914; Hariharānanda Āraṇya, 2000; Bryant, 2009). Read alongside the motor-control evidence, this gives the workshop's slow movement practice its two instructions: use only the effort the posture needs, and keep attention on the sensations of moving.

### *3.5 Feedback*

Each action has consequences: a change in the world, a change in the body, and a change in the nervous system that produced it. The model calls these consequences feedback and places them beneath the four steps rather than after them, because they alter every step at once.

At the cellular level, connections that are used together are strengthened. Hebb (1949) proposed this principle, and it is now summarised as "cells that fire together wire together". Strengthening occurs on both sides of the synapse. On the sending side, repeated activity changes how much transmitter is released per signal (Zucker & Regehr, 2002). On the receiving side, receptor numbers and responsiveness change. Unlike the electrical signal within a neuron, transmission between neurons is chemical. Transmitter is released from vesicles and diffuses across the synapse, a process shaped by vesicle and pore geometry (Kapila & Nair, 2025) and slower than electrical conduction (Kandel et al., 2021). The lasting changes that strengthen a connection unfold more slowly still, over minutes to days. Fast, habitual movement repeats existing connections; slow movement with attention gives the nervous system time and clear feedback with which to build new ones. The reverse also holds: pathways that are not used weaken. Repetition therefore changes the thresholds and timing of every step that took part in an action.

At the behavioural level, the same process produces habits. An action repeated in a stable context becomes linked to that context and is triggered by it with less deliberation (Wood & Rünger, 2016). In the terms of the loop, feedback lowers the threshold at Sense for the familiar cue, biases Process towards the familiar meaning, moves Choose towards reaction, and makes the practised execution easier to run. This is why a pattern feels like "who I am" rather than "what I did".

At the physiological level, repeated activation without recovery has a cost. Allostasis is the process of maintaining stability through change. Allostatic load is the wear that accumulates when stress responses are switched on too often or fail to switch off (McEwen, 1998; Sterling, 2012). Over time the baseline itself drifts, which the autonomic space model in 3.4 describes as a stuck state.

The feedback layer also explains how change happens. If repetition built a pattern, different repetition can replace it with a new pattern that serves better in the present context, as with the examples in 3.3. Each practice in Section 4 aims to produce a different outcome at one step: a safer reading at Process, a pause at Choose, a released muscle at Execute. Repeated, that outcome feeds back and becomes the new default. Because each step feeds the next, a loop can be interrupted at any of them. There is no single correct entry point. Completing an interrupted defensive response, as described in 3.4, is a direct example: SE proposes that the bodily feedback from completion signals that the threat is over (Payne et al., 2015). Memory reconsolidation research offers a related mechanism. Recalled emotional memories can be updated if new experience is added while they are active (Lane et al., 2015).

## 4. Where practices act

The loop's practical claim is that each practice changes one step more than the others, and that repetition then carries the change into the whole loop through feedback. This section places the practices taught in the workshop on the loop. Practices are offered at every step because people differ in where change comes most easily. One person may find it easier to notice body sensations, another to pause, another to move slowly. Each is encouraged to start with the step that is most accessible to them. Table 1 summarises the proposed mechanism and the current strength of evidence for each.

Two dosing rules from SE apply to every practice (Payne et al., 2015). Titration means working in small amounts: a few minutes of attention to a difficult sensation, not an hour. Pendulation means moving deliberately between activation and settling, for example between a tense area and a comfortable one, or between an inner sensation and the view of the room. Both aim to keep the person within the range of arousal they can tolerate while staying aware, often called the window of tolerance (Siegel, 1999). In SE and clinical practice, work outside that window is held to reinforce the threat pattern it was meant to change (Payne et al., 2015). Practices are therefore taught short, frequent and observational. Participants are asked to notice, not to correct.

### *4.1 Practices at Sense*

Full sensory awareness is a guided scan lasting five to seven minutes. Attention moves from the soles of the feet up through the legs, back, arms and breath, then out to sounds and the room, without trying to change anything. It trains attention across all three channels and rehearses noticing without judging, a skill that interoceptive training studies link to better emotion regulation (Price & Hooven, 2018). Grounding focuses on contact and weight: the feet on the floor, the back against the chair. Proprioceptive signals of this kind are steady and easy to detect, which makes them a reliable anchor when interoceptive signals are noisy.

### *4.2 Practices at Process*

Orientation asks the person to look slowly around the room and let the eyes rest on whatever they find pleasing or neutral. It supplies outside evidence of safety. The model proposes that this gives exteroceptive cues more weight relative to bodily alarm. Resourcing asks the person to recall, or notice, something that feels safe or supportive, and to observe what happens in the body. It changes the context in which body signals are read. Reducing stimulus and intensity (fewer notifications, quieter inputs, breaks between demands) lowers the load on processing capacity.

### *4.3 Practices at Choose*

The pause is taught as a brief stop before acting, in which the person names what they are sensing and feeling. Over time the pause is intended to become available in ordinary situations. Simplifying routines and reducing the number of decisions in a day applies the Hick–Hyman law: fewer choices leave more capacity for the decisions that matter.

### *4.4 Practices at Execute and Feedback*

Guided relaxation (recordings of about 3, 7 and 25 minutes) and alternating periods of stimulation and rest train autonomic flexibility. Slow, mindful movement and holding a posture just short of discomfort, without forcing, train antagonist release and fine motor control. Evidence on stretch intensity is mixed. Low-intensity stretching can improve range of motion with less discomfort (Apostolopoulos et al., 2015), while stretching past the pain threshold gives larger short-term gains (Hatano et al., 2022). The workshop aims at lasting change rather than short-term range, so it stays at the gentler end. Pain triggers a protective response that works against relaxation and attention. Practice that stays comfortable is also easier to repeat daily, and repetition is what the feedback layer depends on. Daily repetition is the practice that acts on Feedback. The workshop asks participants to choose one practice each week and repeat it, because only repeated outcomes change the pattern.

**Table 1.** Practices placed on the loop. Evidence strength is the author's judgement of the published literature and is to be refined.

| Practice | Main step | Proposed mechanism | Evidence |
|---|---|---|---|
| Full sensory awareness | Sense | Attention across all three channels; more reliable interoceptive signals | Moderate (interoceptive and body-scan training) |
| Grounding | Sense | Steady proprioceptive input as an anchor | Limited |
| Orientation | Process | Outside cues of safety outweigh bodily alarm | Limited; consistent with safety-signal research |
| Resourcing | Process | A safe context changes the meaning of body signals | Limited |
| Less stimulus and intensity | Process | Lower processing load | Limited for wellbeing |
| The pause | Choose | Shift from habitual to goal-directed control; decentering | Moderate |
| Fewer choices | Choose | Shorter selection time (Hick–Hyman law) | Strong for reaction time; limited for wellbeing |
| Guided relaxation; alternating stimulation and rest | Execute (autonomic) | Autonomic flexibility and recovery | Moderate |
| Slow movement; holding short of discomfort | Execute (somatic) | Antagonist release; finer motor control | Limited to moderate |
| Daily repetition | Feedback | Repeated outcomes rebuild the pattern | Strong for plasticity; limited for these practices |
| Titration and pendulation (dosing) | All steps | Keep practice within the window of tolerance | Limited (SE trials; not tested separately) |

## 5. Propositions

A framework earns its place only if it makes claims that could turn out to be wrong. The loop yields seven propositions. Each names the step it concerns and a validated measure that could test it in non-clinical adults who learn the model.

**P1. Learning the loop increases noticing of reaction versus response.** Adults who learn the four steps will report more occasions on which they notice an automatic reaction before or while it runs. This concerns Choose. It can be tested with a short knowledge check and structured interview items on felt choice.

**P2. Full sensory awareness practice increases interoceptive awareness and trust.** Regular practice will raise scores on noticing, not-distracting, attention regulation and trusting on the Multidimensional Assessment of Interoceptive Awareness (MAIA-2; Mehling et al., 2018). This concerns Sense.

**P3. Orientation and grounding lower state anxiety within a session.** A few minutes of orienting to the room and attending to contact will reduce state anxiety and heart rate compared with an equal period of rest. This concerns Process. It can be tested with the state scale of the State-Trait Anxiety Inventory (Spielberger et al., 1983) and heart rate where available.

**P4. Reducing stimulus and choice load lowers perceived stress.** Participants who apply the four ways of reducing load will report lower perceived stress on the PSS-10 (Cohen et al., 1983). This concerns Process and Choose.

**P5. Slow movement and relaxation practice increase ease of relaxing.** Regular practice will improve self-rated ease of relaxing and the MAIA-2 self-regulation scale. This concerns Execute.

**P6. Effects grow with practice.** Change on P2 to P5 will be larger in participants who log more days of practice. This concerns Feedback and is the most direct test of the feedback layer.

**P7. Reduced stress works through decentering and interoceptive trust.** The fall in perceived stress will be partly mediated by gains in decentering, measured with the Experiences Questionnaire (Fresco et al., 2007), and in MAIA-2 trusting. This concerns the whole loop.

Two of these, P3 and P6, are the most discriminating. P3 can be tested in a single session with an active comparison. P6 distinguishes the model's account of change, which depends on repetition, from a general effect of attending a workshop.

## 6. Relation to existing approaches

The loop borrows from several established approaches. What it adds is one map on which their practices can be placed, taught to non-clinical adults as an account of their own nervous system.

**Somatic Experiencing.** SE is the closest relative. It supplies the bottom-up order of attention, resourcing and orienting, titration and pendulation, and the idea of completing interrupted defensive responses (Payne et al., 2015). The differences are of setting and aim. SE is a one-to-one trauma therapy delivered by trained practitioners. The loop is taught in groups to adults without a clinical diagnosis, and it explains the mechanism so that people can apply the practices themselves. SE's evidence base is still preliminary (Kuhfuß et al., 2021), so the loop uses its concepts as hypotheses, not as established findings.

**Pain neuroscience education.** Pain neuroscience education teaches people with chronic pain how the nervous system produces pain, and has been found to reduce pain catastrophising and fear of movement (Moseley & Butler, 2015). It shows that understanding the mechanism can itself change experience. The loop applies the same logic to everyday stress, a domain that has no comparable teaching model.

**Mindfulness and interoceptive training.** Mindfulness-based programmes train non-judgmental attention and decentering (Bishop et al., 2004). Mindful awareness in body-oriented therapy trains interoceptive skills for emotion regulation (Price & Hooven, 2018). The loop shares their practices at Sense and Choose. It differs by naming where each practice acts and by adding practices at Execute and Feedback, such as slow movement and planned repetition, that these programmes treat as secondary.

**Acceptance and commitment therapy.** ACT targets psychological flexibility through acceptance, defusion and committed action (Hayes et al., 2006). The pause at Choose resembles defusion, and the loop's emphasis on flexibility over constant calm is close to ACT's. The loop gives more weight to bodily mechanisms and less to values and language.

**Coaching.** In coaching, the loop gives coach and client a shared language for what happens between a trigger and a behaviour. It also helps the coach recognise when a client's reaction is driven by bodily state rather than by a lack of insight.

**Yoga.** Yoga therapy is the third source of the model, alongside SE and coaching. Several workshop practices come from it: posture held with relaxed effort and attention (3.4), breath observation, and guided relaxation. Yoga-based interventions lower some physiological markers of stress (Pascoe et al., 2017), and their self-regulatory effects have been described in terms of top-down attention and bottom-up bodily signals (Gard et al., 2014). The loop differs in vocabulary, not aim: it explains in neuroscience terms why these practices work, and places each on a step.

**From self-regulation to practice.** The loop has been translated into a five-session certificate course for coaches and other helping professionals. Participants watch the matching workshop session in advance and then practise one skill per session in small groups. Table 2 places the four skills on the loop. Presence draws on the literature on therapeutic presence (Geller & Greenberg, 2012). An evaluation of the course will be reported separately.

**Table 2.** Practitioner skills placed on the loop.

| Skill | Main step | Basis |
|---|---|---|
| Non-judgmental, compassionate presence that offers the client cues of safety | Process (client); whole loop (practitioner) | Therapeutic presence; co-regulation |
| Full sensory awareness with titration and pendulation | Sense, Process | Interoceptive training; SE dosing; window of tolerance |
| Completing interrupted defensive responses, with co-regulation and education | Execute, Feedback | SE; defence cascade; not tested directly |
| Identifying and weakening patterns that create loops and overwhelm, with curiosity and stillness | Feedback, Choose | Memory reconsolidation (Lane et al., 2015); decentering |

## 7. Safety, scope and limits

The loop is a framework for wellbeing and self-regulation in adults without a clinical diagnosis. It is not a treatment, and it is not a substitute for care for trauma, anxiety disorders, depression or physical illness.

Body-based practices are not risk-free. Attention to internal sensation can bring up strong emotion, memories or bodily responses such as shaking, tears or dizziness, especially in people with a history of trauma. Three safeguards follow from the model itself. Practices are dosed by titration and pendulation and kept within the person's window of tolerance. Participants are invited to notice, not required to stay with discomfort, and can return attention to the room at any time. People who report significant distress are referred to an appropriate professional. Group delivery adds a fourth safeguard: facilitators should screen for current crisis or recent trauma before enrolment.

The model has four limits. First, it is a synthesis, not a new finding. Its parts rest on evidence of very different strength, from the well-replicated Hick–Hyman law to SE's hypothesis about completing defensive responses (Table 1). Second, the division into four steps is a teaching simplification. In the brain these processes overlap, run in parallel and influence one another continuously, and predictive accounts would place feedback within every step as well as beneath them. Third, the model was developed by its teacher, and practitioner observations are open to confirmation bias. Fourth, there is as yet no outcome data. The propositions in Section 5 are meant to be tested, and a retrospective and a prospective evaluation of the workshop are planned.

## 8. Conclusion

The Four-Step Action Loop with Feedback offers adults a map of their own stress responses. A cue is sensed, given meaning, met with a reaction or a response, and carried out by muscles and organs. Each outcome then feeds back to shape every step the next time. The model adds no new mechanism. Its value lies in arranging established neuroscience so that a non-specialist can locate their experience on it and see why a given practice should help. If the propositions hold, the loop could do for everyday stress what pain neuroscience education has done for pain: turn understanding of the mechanism into a means of change.

**Declaration of generative AI use**

The model, practices and source material are the author's own, developed through the author’s workshop teaching. During the preparation of this work the author used Claude (Anthropic) to help draft and edit text from author notes and content for the workshop, and check references. After using this tool, the author reviewed and edited the content as needed and takes full responsibility for the content of the published article.

**Declaration of competing interest**

The author is co-founder of Swasth Foundation, whose programme The Swasth Institute offers the workshop and certificate course described in this paper for a fee, with scholarships.

**Funding**

This research did not receive any specific grant from funding agencies in the public, commercial, or not-for-profit sectors.

**Acknowledgements**

The author's greatest teachers have been the workshop participants and clients whose experiences and questions shaped this framework, and he is deeply grateful to them. He also thanks his teachers in Somatic Experiencing, yoga and coaching, and his colleagues at The Swasth Institute and Swasth Foundation for their support.

**CRediT authorship contribution statement**

Sundeep Kapila: Conceptualization, Methodology, Writing – original draft, Writing – review & editing.

**Data availability**

No data were used for the research described in the article.

## References

Andersen, T. E., Lahav, Y., Ellegaard, H., & Manniche, C. (2017). A randomized controlled trial of brief somatic experiencing for chronic low back pain and comorbid post-traumatic stress disorder symptoms. *European Journal of Psychotraumatology, 8*(1), 1331108. https://doi.org/10.1080/20008198.2017.1331108

Apostolopoulos, N., Metsios, G. S., Flouris, A. D., Koutedakis, Y., & Wyon, M. A. (2015). The relevance of stretch intensity and position: A systematic review. *Frontiers in Psychology, 6*, 1128. https://doi.org/10.3389/fpsyg.2015.01128

Balar, P., & Kapila, S. (2025). *Integrative model for interoception and exteroception: Predictive coding, points of modulation, and testable predictions* (arXiv:2511.13668). arXiv. https://doi.org/10.48550/arXiv.2511.13668

Barrett, L. F., & Simmons, W. K. (2015). Interoceptive predictions in the brain. *Nature Reviews Neuroscience, 16*(7), 419–429. https://doi.org/10.1038/nrn3950

Bernstein, A., Hadash, Y., Lichtash, Y., Tanay, G., Shepherd, K., & Fresco, D. M. (2015). Decentering and related constructs: A critical review and metacognitive processes model. *Perspectives on Psychological Science, 10*(5), 599–617. https://doi.org/10.1177/1745691615594577

Berntson, G. G., Cacioppo, J. T., & Quigley, K. S. (1991). Autonomic determinism: The modes of autonomic control, the doctrine of autonomic space, and the laws of autonomic constraint. *Psychological Review, 98*(4), 459–487. https://doi.org/10.1037/0033-295X.98.4.459

Bishop, S. R., Lau, M., Shapiro, S., Carlson, L., Anderson, N. D., Carmody, J., Segal, Z. V., Abbey, S., Speca, M., Velting, D., & Devins, G. (2004). Mindfulness: A proposed operational definition. *Clinical Psychology: Science and Practice, 11*(3), 230–241. https://doi.org/10.1093/clipsy.bph077

Brom, D., Stokar, Y., Lawi, C., Nuriel-Porat, V., Ziv, Y., Lerner, K., & Ross, G. (2017). Somatic experiencing for posttraumatic stress disorder: A randomized controlled outcome study. *Journal of Traumatic Stress, 30*(3), 304–312. https://doi.org/10.1002/jts.22189

Brosschot, J. F., Verkuil, B., & Thayer, J. F. (2016). The default response to uncertainty and the importance of perceived safety in anxiety and stress: An evolution-theoretical perspective. *Journal of Anxiety Disorders, 41*, 22–34. https://doi.org/10.1016/j.janxdis.2016.04.012

Brosschot, J. F., Verkuil, B., & Thayer, J. F. (2018). Generalized unsafety theory of stress: Unsafe environments and conditions, and the default stress response. *International Journal of Environmental Research and Public Health, 15*(3), 464. https://doi.org/10.3390/ijerph15030464

Bryant, E. F. (2009). *The Yoga Sūtras of Patañjali: A new edition, translation, and commentary with insights from the traditional commentators*. North Point Press.

Cohen, S., Kamarck, T., & Mermelstein, R. (1983). A global measure of perceived stress. *Journal of Health and Social Behavior, 24*(4), 385–396. https://doi.org/10.2307/2136404

Craig, A. D. (2002). How do you feel? Interoception: The sense of the physiological condition of the body. *Nature Reviews Neuroscience, 3*(8), 655–666. https://doi.org/10.1038/nrn894

Dolan, R. J., & Dayan, P. (2013). Goals and habits in the brain. *Neuron, 80*(2), 312–325. https://doi.org/10.1016/j.neuron.2013.09.007

Fresco, D. M., Moore, M. T., van Dulmen, M. H. M., Segal, Z. V., Ma, S. H., Teasdale, J. D., & Williams, J. M. G. (2007). Initial psychometric properties of the Experiences

Questionnaire: Validation of a self-report measure of decentering. *Behavior Therapy, 38*(3), 234–246. https://doi.org/10.1016/j.beth.2006.08.003

Friston, K. (2010). The free-energy principle: A unified brain theory? *Nature Reviews Neuroscience, 11*(2), 127–138. https://doi.org/10.1038/nrn2787

Fuster, J. M. (2004). Upper processing stages of the perception–action cycle. *Trends in Cognitive Sciences, 8*(4), 143–145. https://doi.org/10.1016/j.tics.2004.02.004

Gard, T., Noggle, J. J., Park, C. L., Vago, D. R., & Wilson, A. (2014). Potential self-regulatory mechanisms of yoga for psychological health. *Frontiers in Human Neuroscience, 8*, 770. https://doi.org/10.3389/fnhum.2014.00770

Garfinkel, S. N., & Critchley, H. D. (2016). Threat and the body: How the heart supports fear processing. *Trends in Cognitive Sciences, 20*(1), 34–46. https://doi.org/10.1016/j.tics.2015.10.005

Garfinkel, S. N., Seth, A. K., Barrett, A. B., Suzuki, K., & Critchley, H. D. (2015). Knowing your own heart: Distinguishing interoceptive accuracy from interoceptive awareness. *Biological Psychology, 104*, 65–74. https://doi.org/10.1016/j.biopsycho.2014.11.004

Geller, S. M., & Greenberg, L. S. (2012). *Therapeutic presence: A mindful approach to effective therapy*. American Psychological Association. https://doi.org/10.1037/13485-000

Hariharānanda Āraṇya. (2000). *Yoga philosophy of Patañjali with Bhāsvatī* (4th enl. ed.). University of Calcutta.

Hatano, G., Matsuo, S., Asai, Y., Suzuki, S., & Iwata, M. (2022). Effects of high-intensity stretch with moderate pain and maximal intensity stretch without pain on flexibility. *Journal of Sports Science and Medicine, 21*(2), 171–181. https://doi.org/10.52082/jssm.2022.171

Hayes, S. C., Luoma, J. B., Bond, F. W., Masuda, A., & Lillis, J. (2006). Acceptance and commitment therapy: Model, processes and outcomes. *Behaviour Research and Therapy, 44*(1), 1–25. https://doi.org/10.1016/j.brat.2005.06.006

Hebb, D. O. (1949). *The organization of behavior: A neuropsychological theory*. Wiley.

Henneman, E. (1957). Relation between size of neurons and their susceptibility to discharge. *Science, 126*(3287), 1345–1347. https://doi.org/10.1126/science.126.3287.1345

Henneman, E., Somjen, G., & Carpenter, D. O. (1965). Functional significance of cell size in spinal motoneurons. *Journal of Neurophysiology, 28*(3), 560–580. https://doi.org/10.1152/jn.1965.28.3.560

Hick, W. E. (1952). On the rate of gain of information. *Quarterly Journal of Experimental Psychology, 4*(1), 11–26. https://doi.org/10.1080/17470215208416600

Hillier, S., & Worley, A. (2015). The effectiveness of the Feldenkrais method: A systematic review of the evidence. *Evidence-Based Complementary and Alternative Medicine, 2015*, 752160. https://doi.org/10.1155/2015/752160

Hyman, R. (1953). Stimulus information as a determinant of reaction time. *Journal of Experimental Psychology, 45*(3), 188–196. https://doi.org/10.1037/h0056940

Jaakkola, E. (2020). Designing conceptual articles: Four approaches. *AMS Review, 10*(1–2), 18–26. https://doi.org/10.1007/s13162-020-00161-0

Kandel, E. R., Koester, J. D., Mack, S. H., & Siegelbaum, S. A. (Eds.). (2021). *Principles of neural science* (6th ed.). McGraw Hill.

Kapila, S., & Nair, P. R. (2025). Extraction of geometric and transport parameters from the time constant of exocytosis transients measured by nanoscale electrodes. *European Biophysics Journal, 54*(3), 149–157. https://doi.org/10.1007/s00249-025-01744-0

Khalsa, S. S., Adolphs, R., Cameron, O. G., Critchley, H. D., Davenport, P. W., Feinstein, J. S., Feusner, J. D., Garfinkel, S. N., Lane, R. D., Mehling, W. E., Meuret, A. E., Nemeroff, C. B., Oppenheimer, S., Petzschner, F. H., Pollatos, O., Rhudy, J. L., Schramm, L. P., Simmons, W. K., Stein, M. B., … Zucker, N. (2018). Interoception and mental health: A roadmap. *Biological Psychiatry: Cognitive Neuroscience and Neuroimaging, 3*(6), 501–513. https://doi.org/10.1016/j.bpsc.2017.12.004

Kozlowska, K., Walker, P., McLean, L., & Carrive, P. (2015). Fear and the defense cascade: Clinical implications and management. *Harvard Review of Psychiatry, 23*(4), 263–287. https://doi.org/10.1097/HRP.0000000000000065

Kuhfuß, M., Maldei, T., Hetmanek, A., & Baumann, N. (2021). Somatic experiencing – effectiveness and key factors of a body-oriented trauma therapy: A scoping literature review. *European Journal of Psychotraumatology, 12*(1), 1929023. https://doi.org/10.1080/20008198.2021.1929023

Lane, R. D., Ryan, L., Nadel, L., & Greenberg, L. (2015). Memory reconsolidation, emotional arousal, and the process of change in psychotherapy: New insights from brain science. *Behavioral and Brain Sciences, 38*, e1. https://doi.org/10.1017/S0140525X14000041

McEwen, B. S. (1998). Protective and damaging effects of stress mediators. *New England Journal of Medicine, 338*(3), 171–179. https://doi.org/10.1056/NEJM199801153380307

Mehling, W. E., Acree, M., Stewart, A., Silas, J., & Jones, A. (2018). The Multidimensional Assessment of Interoceptive Awareness, Version 2 (MAIA-2). *PLoS ONE, 13*(12), e0208034. https://doi.org/10.1371/journal.pone.0208034

Merkel, J. (1885). Die zeitlichen Verhältnisse der Willensthätigkeit. *Philosophische Studien, 2*, 73–127.

Moseley, G. L., & Butler, D. S. (2015). Fifteen years of explaining pain: The past, present, and future. *Journal of Pain, 16*(9), 807–813. https://doi.org/10.1016/j.jpain.2015.05.005

Pascoe, M. C., Thompson, D. R., & Ski, C. F. (2017). Yoga, mindfulness-based stress reduction and stress-related physiological measures: A meta-analysis. *Psychoneuroendocrinology, 86*, 152–168. https://doi.org/10.1016/j.psyneuen.2017.08.008

Paulus, M. P., & Stein, M. B. (2010). Interoception in anxiety and depression. *Brain Structure and Function, 214*(5–6), 451–463. https://doi.org/10.1007/s00429-010-0258-9

Payne, P., Levine, P. A., & Crane-Godreau, M. A. (2015). Somatic experiencing: Using interoception and proprioception as core elements of trauma therapy. *Frontiers in Psychology, 6*, 93. https://doi.org/10.3389/fpsyg.2015.00093

Price, C. J., & Hooven, C. (2018). Interoceptive awareness skills for emotion regulation: Theory and approach of mindful awareness in body-oriented therapy (MABT). *Frontiers in Psychology, 9*, 798. https://doi.org/10.3389/fpsyg.2018.00798

Schmidt, R. A., Lee, T. D., Winstein, C. J., Wulf, G., & Zelaznik, H. N. (2019). *Motor control and learning: A behavioral emphasis* (6th ed.). Human Kinetics.

Schwabe, L., & Wolf, O. T. (2011). Stress-induced modulation of instrumental behavior: From goal-directed to habitual control of action. *Behavioural Brain Research, 219*(2), 321–328. https://doi.org/10.1016/j.bbr.2010.12.038

Seth, A. K. (2013). Interoceptive inference, emotion, and the embodied self. *Trends in Cognitive Sciences, 17*(11), 565–573. https://doi.org/10.1016/j.tics.2013.09.007

Siegel, D. J. (1999). *The developing mind: Toward a neurobiology of interpersonal experience*. Guilford Press.

Spielberger, C. D., Gorsuch, R. L., Lushene, R., Vagg, P. R., & Jacobs, G. A. (1983). *Manual for the State-Trait Anxiety Inventory (Form Y)*. Consulting Psychologists Press.

Sterling, P. (2012). Allostasis: A model of predictive regulation. *Physiology & Behavior, 106*(1), 5–15. https://doi.org/10.1016/j.physbeh.2011.06.004

Thayer, J. F., & Lane, R. D. (2000). A model of neurovisceral integration in emotion regulation and dysregulation. *Journal of Affective Disorders, 61*(3), 201–216. https://doi.org/10.1016/S0165-0327(00)00338-4

Wood, W., & Rünger, D. (2016). Psychology of habit. *Annual Review of Psychology, 67*, 289–314. https://doi.org/10.1146/annurev-psych-122414-033417

Woods, J. H. (1914). *The Yoga-system of Patañjali* (Harvard Oriental Series, Vol. 17). Harvard University Press.

Zucker, R. S., & Regehr, W. G. (2002). Short-term synaptic plasticity. *Annual Review of Physiology, 64*, 355–405. https://doi.org/10.1146/annurev.physiol.64.092501.114547